\documentclass[aps,prd,reprint]{revtex4-2}

\usepackage[T1]{fontenc}
\usepackage[utf8]{inputenc}
\usepackage{amsfonts,amsmath,amssymb,accents,mathrsfs}
\usepackage[retainorgcmds]{IEEEtrantools}
\usepackage{tikz}
\usepackage{graphicx,color}
\usepackage{booktabs}
\usepackage{placeins}
\usepackage[colorlinks,
			breaklinks,
			hyperfootnotes,
			linktoc=page,
            linkcolor=blue,
            citecolor=blue,
            urlcolor=blue,
            unicode,
            pdfstartview=FitH
            bookmarks,
            bookmarksnumbered]{hyperref}

\renewcommand\a {{\alpha}}
\renewcommand\b {{\beta}}
\newcommand\g {{\gamma}}
\renewcommand\d {{\delta}}
\renewcommand\r {{\rho}}

\renewcommand\L {{\Lambda}}

\newcommand\ad {{\dot{\alpha}}}
\newcommand\bd {{\dot{\beta}}}
\newcommand\gd {{\dot{\gamma}}}

\newcommand\rd {{\dot{\rho}}}

\newcommand\D {{\rm D}}
\newcommand\Dd {{\bar{\rm D}}}
\newcommand\pa {{\partial}}

\newcommand\seqn[1] {\begin{equation}#1\end{equation}}

\def\bea{\begin{IEEEeqnarray*}}
\def\eea{\end{IEEEeqnarray*}}
\def\be{\begin{eqnarray}}
\def\ee{\end{eqnarray}}
\def\n{\IEEEyesnumber}
\def\sn{\IEEEyessubnumber}

\newcommand\nn {\nonumber}
\begin{document}

\title{Hierarchy of Supersymmetric Higher Spin Connections}


\author{I.~L.~Buchbinder}
\email{joseph@tspu.edu.ru}
\affiliation{Department of Theoretical Physics, Tomsk State Pedagogical University, Tomsk, 634061, Russia}
\affiliation{National Research Tomsk State University, Tomsk 634050, Russia}
\author{S.~James~Gates~Jr.}
\email{sylvester\_gates@brown.edu}
\author{K.~Koutrolikos}
\email{konstantinos\_koutrolikos@brown.edu}
\affiliation{Brown Theoretical Physics Center,
      Box S, 340 Brook Street, Barus Hall,
      Providence, RI 02912, USA}
\affiliation{Department of Physics, Brown University,
      Box 1843, 182 Hope Street, Barus \& Holley 545,
  Providence, RI 02912, USA}


\begin{abstract}
    We focus on the geometrical reformulation of free higher spin supermultiplets in
    $4\D,~\mathcal{N}=1$ flat superspace. We find that there is a de Wit-Freedman like hierarchy of
    superconnections with simple gauge transformations. The requirement for sensible free equations
    of motion imposes constraints on the gauge parameter superfields. Unlike the nonsupersymmetric
    case, we find several different constraints that can decouple the higher superconnections.
    By lifting these constraints nongeometrically via compensators we recover all known descriptions
    of arbitrary  integer and half-integer gauge supermultiplets. In the constrained formulation we find
    a new description of half-integer supermultiplets, generalizing the new-minimal and virial
    formulations of linearized supergravity to higher spins. However this description can be formulated
    using compensators. The various descriptions can be labeled as
    geometrical or nongeometrical if the equations of motion can be expressed purely in terms of
    superconnections or not.
\end{abstract}


\maketitle

\section{Introduction}\label{sec:intro}
The study of higher spins plays a special role in the search for underlying principles and
symmetries of nature. Depending on your viewpoint this statement can be understood in two
different ways. On the one hand, their contribution to string theory is crucial~\cite{Isberg:1993av,
Sundborg:2000wp,Sezgin:2002rt,Sagnotti:2003qa,Engquist:2005yt,Bonelli:2003kh,Fotopoulos:2007nm} and
higher spin symmetry may control the UV completion of gravity~\cite{Fradkin:1986ka,Gross:1988ue}.
On the other hand, various no-go results~\cite{Weinberg:1964ew,Weinberg:1980kq,Bekaert:2010hw,Rahman:2015pzl,
Taronna:2017wbx} under specific assumptions constrain the list of nontrivial interactions among particles.

Most of the progress done in higher-spin theories falls under two categories: (\emph{i})
constructing consistent interactions involving higher spin gauge fields
and (\emph{ii}) the geometrical reformulation of free higher spins on Minkowski and AdS
backgrounds.  The correlation between these two directions is evident in the case of gravity,
where the geometrical formulation of the theory dictates its interactions. By analogy,
understanding the underlying geometrical structure of higher spin theory, assuming one exist,
may provide a deeper insight to higher spin interactions.

Nontrivial higher spin interactions have been constructed employing a variety of techniques such as
Noether method~\cite{Berends:1985xx,Barnich:1993vg,Bekaert:2009ud,Bekaert:2010hk,Buchbinder:2012iz,
Joung:2012fv,Joung:2013nma},
BRST~\cite{Buchbinder:2006eq,Bekaert:2006us,Fotopoulos:2007yq,Fotopoulos:2008ka,Polyakov:2009pk,
Henneaux:2012wg,Metsaev:2012uy,Henneaux:2013gba}, light
cone~\cite{Bengtsson:1983pd,Metsaev:2005ar,Metsaev:2007rn,Metsaev:2017cuz},
and framelike formulation~\cite{Vasiliev:1980as,Fradkin:1987ks,Lopatin:1987hz,Vasiliev:1987tk,
Vasiliev:2001wa,Alkalaev:2002rq,Zinoviev:2008ze,Ponomarev:2010st,Zinoviev:2010cr,
Boulanger:2011qt,Zinoviev:2011fv,Buchbinder:2019dof,Buchbinder:2019olk,Buchbinder:2019kuh,
Buchbinder:2020rex,Khabarov:2020bgr}.
The framelike description has been the most successful and provides an economy of ideas \footnote{It
resembles the structure of a non-abelian Yang-Mills theory for the
group of isometries of the underlying manifold and its higher spin symmetry extension.}. However,
the metriclike approach
offers an economy of fields which makes the geometrical interpretation of the theory more direct \footnote{
The seemingly disconnected choices of approach (a)
framelike (gauging of the underlying symmetry group) or (b) metriclike (geometry) align
coherently in Felix Klein's view of geometry as the action of a Lie group $G$ on coset spaces
$G/H$~\cite{klein2008comparative} and Cartan's generalization of it~\cite{sharpe1997cartans}.}.
This was first demonstrated by de Wit and Freedman in~\cite{deWit:1979sib}. They found that for a
bosonic spin $s$, the object replacing the usual connection of Riemannian geometry
is a tower of $s-1$ connectionlike
objects \footnote{For the case of gravity, $s=1$, this tower collapses to one connection, the
Christoffel symbol.}, each being the derivative of the previous one. The top connection allowed
the definition of an invariant curvature tensor which is the $s$-th spacetime derivative of
the higher spin gauge field. Extracting Fr\o{}nsdal's second order equation of motion required
imposing a traceless constraint on the gauge parameter in
order to decouple the higher order connections \footnote{Higher connectionlike objects also appear
in the framelike description of higher spins~\cite{Vasiliev:1986td,Fradkin:1986qy,Vasiliev:2001wa}.
These are called `auxiliary fields' but they can be decoupled from the free theory by traceless
conditions, precisely to get two derivative equations.}, thus reducing the symmetry group.
Later an unconstrained, geometrical but nonlocal description~\cite{Francia:2002aa,
Francia:2002pt} was found together with an  unconstrained, local but nongeometrical
description~\cite{Francia:2005bu,Francia:2007qt}.

Higher spin theories with manifest $\mathcal{N}=1$ supersymmetry were first formulated
in~\cite{Kuzenko:1993jp, Kuzenko:1993jq,Kuzenko:1994dm} and later works
include~\cite{Gates:1996my,Gates:2013rka,Gates:2013ska,Gates:2013tka,Kuzenko:2016bnv,
Kuzenko:2016qwo,Kuzenko:2017ujh,Buchbinder:2019esz}). Recently
there has been some progress in the direction of constructing consistent
interactions~\cite{Buchbinder:2017nuc,Hutomo:2017phh,Hutomo:2017nce,Koutrolikos:2017qkx,
Buchbinder:2018wwg,Buchbinder:2018wzq,Buchbinder:2018nkp,Buchbinder:2018gle,Hutomo:2018tjh,Gates:2019cnl,Metsaev:2019dqt,Metsaev:2019aig}
for such theories.
Nevertheless, no steps have been taken towards the geometrical reformulation of these theories.  This paper
is a first analysis in that direction.

We study the properties of a set of natural objects which define the notion of generalized higher spin
superconnection and the corresponding  supercurvature superfield.
We find that these objects arrange into a hierarchy \`a la de Wit and Freedman~\cite{deWit:1979sib} in
superspace. The top member of this hierarchy is a proper superconnection in the sense that it
allows the definition of an invariant supercurvature superfield.
These supercurvatures match the known higher spin superfield strengths.
Demanding sensible superspace equations of motion for free theory, generates a variety of nonequivalent
constraints that one can impose on the gauge parameter superfield. This is in contrast to
nonsupersymmetric theories, where there is a unique constraint, the traceless condition of the  parameter.
In superspace, we find that there is more than one ways one can decouple the higher order superconnections
and that gives rise to all these different constraints. Breaking our geometrical approach, we use
the method of compensators to lift the constraints and show that all known descriptions of higher and
lower spin supermultiplets correspond to one of such constraints. However, for one of the constraints this
can not be done. For that case, the constrained formulation gives a new description of the
half integer superspin supermultiplet $(s+1,s+1/2)$. This new description is a higher spin generalization
of the known new-minimal and new-new-minimal descriptions of linearized supergravity.
Finally, we find that in the constrained formulation some of the theories have the property
that their equation of motion can be written purely in terms of superconnections. This property gives
a sense of geometrical origin for these theories.

The paper is organized as follows. In section \ref{sec:superconnections}, getting inspiration from
super Yang-Mills theory, we define the notion of a generalized superconnection and its
corresponding supercurvature tensor. In section \ref{sec:hidWFs}, we focus on the $(s+1,s+1/2)$ class of
supermultiplets which are described by a bosonic gauge superfield and show that there is a hierarchy
of $(s+1)$ superconnectionlike objects, which at component level contains the known
de Wit-Freedman connections. In section \ref{subsec:eom}, we present
standard equations of motions by constraining the gauge parameter.
In section \ref{subsec:uncformulation}, we introduce nongeometrical compensators
in order to lift constraints and compare with known results.
In section \ref{sec:idWFs}, the
analysis is repeated for the $(s+1/2,s)$ class of supermultiplets described by a fermionic gauge
superfield. In this case there are two independent hierarchies, with $s$ and $s+1$ members
respectively and use them to generate all appropriate constraints in order to extract free equations
of motion.
\section{Superconnections}\label{sec:superconnections}
Gauge redundancy has been proven crucial in constructing manifestly supersymmetric field theories
for higher spins and a particular set of their interactions. However, as it stands the formulation
used in these constructions is not very geometrical. This is because the superspace actions~\cite{
Kuzenko:1993jp,Kuzenko:1993jq,Kuzenko:1994dm,Gates:2013rka,Gates:2013ska} for
free integer and half-integer superspins have been determined by hand and there is no obvious way
of rewriting them in terms of higher spin superfield strengths that involve higher derivatives. A
step towards a more geometrical description would require a generalization of the notion of
superconnection, in the context of higher spins.

Following de Wit and Freedman~\cite{deWit:1979sib}, the signal of a proper
connection is its ability to allow the definition of a gauge invariant tensor in terms of its
derivatives. To identify this signal in manifestly supersymmetric theories and set the stage for
later examinations, let us review the results~\cite{Ferrara:1974pu} regarding super Yang-Mills theory.
For a $U(1)$ gauge group \footnote{Similarly, one can consider
internal symmetries associated to a compact, non-abelian Lie group $G\otimes{U(1)}^p$, where G is
semi-simple.}, the theory is described by a real scalar gauge superfield $V(z)$
with the following gauge transformation:
$e^{gV}\to{e}^{i\bar{\L}}e^{gV}e^{-i\L}\overset{U{(1)}}{\Rightarrow}
\d V=\tfrac{1}{g}\left(\Dd^2 L +\D^2\bar{L}\right)$
where $L$ is the propoetential of chiral superfield $\L$ and is unconstrained.
The covariant derivatives $\nabla_{A}=\{\nabla_{\a},\nabla_{\ad},\nabla_{\a\ad}\}$ are consistent with the
gauge transformation of matter (chiral) superfields that couple to $V$:
$\nabla_A\to{e}^{i\L}\nabla_{A}e^{-i\L}$.
A consistent and convenient choice of covariant derivatives is the following
\seqn{\nabla_{\a}=e^{-gV}\D_{\a}e^{gV},~\nabla_{\ad}=\Dd_{\ad},~\nabla_{\a\ad}=-i\{\nabla_{\a},\nabla_{\ad}\}.}
Their algebra \footnote{We use \emph{Superspace}'s~\cite{Gates:1983nr} conventions.} gives the
corresponding supercurvatures
$W_{\a}=\Dd^2\left(e^{-gV}\D_{\a}e^{gV}\right)$ and
$\bar{W}_{\ad}=e^{-gV}\left[\D^2\left(e^{gV}\Dd_{\ad}e^{-gV}\right)\right]e^{gV}\overset{U(1)}{=}
\D^2\left(e^{gV}\Dd_{\ad}e^{-gV}\right)$.
The superconnections $\Gamma_{A}$ are defined as the
difference between the $U(1)$
covariant derivatives $\nabla_A$ and the supersymmetry covariant derivatives $\D_A$: ($\nabla_A=\D_A+\Gamma_A$)
\seqn{\Gamma_{\a}=e^{-gV}(\D_{\a}e^{gV}),~\Gamma_{\ad}=0,~\Gamma_{\a\ad}=-i\Dd_{\ad}\Gamma_{\a}}
and the supercurvature can be written as $W_{\a}=\Dd^2\Gamma_{\a}$.
For the linearized theory we get the following:
{\small
\bea{l}
\Gamma_{\a}=g\D_{\a}V,~\Gamma_{\ad}=0,~\Gamma_{\a\ad}=-ig\Dd_{\ad}\D_{\a}V,~W_{\a}=\Dd^2\Gamma_{\a}\n
\\[2mm]
\d\Gamma_{\a}=\D_{\a}\Dd^2 L,~\d\Gamma_{\ad}=0,~\d\Gamma_{\a\ad}=\pa_{\a\ad}\Dd^2 L,~\d W_{\a}=0.\n
\eea
}
Projecting \footnote{Denoted by $\Big\rvert$ which is shorthand notation for evaluation at
$\theta=\bar{\theta}=0$.}
the above to components we find (in W.Z gauge):
$\Gamma_{\a}\big\rvert_{W.Z}=0$,~$\Gamma_{\ad}\big\rvert=0$,~$\Gamma_{\a\ad}\big\rvert_{W.Z.}=i~A_{\a\ad}$
~~($\d A_{\a\ad}=\pa_{\a\ad}\xi$)
which is exactly what is expected from field theory.

In the above, we recognize the role of $\Gamma_{\a}$ as a proper connection.
The properties that give it this characterization are (\emph{i})
its transformation has the structure $\D_{\a}\Dd^2 L$ and (\emph{ii})
it allows the definition of an invariant field strength $W_{\a}$ by acting with $\Dd^2$ on it.
Getting inspiration from the above we define a notion of a generalized superconnection in the following
way. In the context of a linearized theory we will call a superfield to be a \emph{superconnection}
if it has a gauge transformation of the form
\seqn{\label{scn}\d\Gamma_{\a\dots}=\D_{\a}\Dd^2\left(\dots\right)}
and therefore it allows the definition of an invariant \emph{supercurvature}
\seqn{\label{scw}W_{\a\dots}=\Dd^2\Gamma_{\a\dots}~,~\d W_{\a\dots}=0~.}
\section{Hierarchy of de Wit-Freedman superconnections for half integer superspins}\label{sec:hidWFs}
Consider a supersymmetric, irreducible system of massless higher spins, in 4D Minkowski spacetime. This
system will include a bosonic and a fermionic higher spin gauge field, which are related by
supersymmetry transformations, hence their spin values must differ by $1/2$. There are two cases,
either the fermion is at the bottom $(s+1,s+1/2)$ [half-integer superspin supermultiplet] or the boson is at the
bottom $(s+1/2,s)$ [integer superspin supermultiplet].

In this section we focus on the half-integer supermultiplet where the
highest propagating spin is $(s+1)$. The appropriate superfield for the description of such a
supermultiplet is a real bosonic $(s,s)$-superfield tensor \footnote{We are using two component
notation, where a single spacetime index is converted to a pair of spinorial indices. In 4D
with Lorentzian signature, spinors can be split to Weyl spinors which have a definite helicity
and carry undotted / dotted indices respectively that take two values. Irreducible $(p,q)$-superfield
tensors carry $p$ undotted indices which are symmetrized (denoted as $\a(p)$) and $q$ dotted indices
which are also independently symmetrized (denoted as $\ad(q)$). Symmetrization of indices with
weight one is denoted by $()$.}
$H_{\a(s)\ad(s)}$. Its highest rank component \footnote{A detailed discussion
regarding the components of a $4D,N=1$ superfield can be found in~\cite{Gates:2017hmb}.} field
is a symmetric $(s+1)$-rank spacetime tensor which will play the role of the highest spin boson
$h_{\a(s+1)\ad(s+1)}\propto
\tfrac{1}{(s+1)!^2}[\D_{(\a_{s+1}},\Dd_{(\ad_{s+1}}]H_{\a(s))\ad(s))}\big\rvert$.
It is easy to verify that
the most general transformation of $H_{\a(s)\ad(s)}$ that gives $h_{\a(s+1)\ad(s+1)}$ the correct
gauge transformation $(\d h_{\a(s+1)\ad(s+1)}\propto \pa_{(\a_{s+1}(\ad_{s+1}}\xi_{\a(s))\ad(s))})$
and is consistent with its reality is:
\seqn{\label{dH}
\d
H_{\a(s)\ad(s)}=\frac{1}{s!}\D_{(\a_s}\bar{L}_{\a(s-1))\ad(s)}-\frac{1}{s!}\Dd_{(\ad_s}L_{\a(s)\ad(s-1))}.
}
Looking back to the super Yang-Mills example, the goal is starting from $H$-superfield and by acting with
spinorial covariant derivatives, to construct a set of objects with simple transformations
under \eqref{dH}. Consider the following quantities \footnote{Because
of the reality of $H$ the objects generated by the exchange of $\D$ and $\Dd$ are not independent.}:
\bea{ll}
\Gamma_{\b\a(s)\ad(s)}&=\D_{\b}H_{\a(s)\ad(s)}~,\n\\
\d\Gamma_{\b\a(s)\ad(s)}&=
-\frac{1}{s!}C_{\b(\a_s}\D^2\bar{L}_{\a(s-1))\ad(s)}\sn\\
&~~-\frac{1}{s!}\D_{\b}\Dd_{(\ad_s}L_{\a(s)\ad(s-1))}~,
\\[4mm]
\Gamma_{\b\a(s)\bd\ad(s)}&=\Dd_{\bd}\D_{\b}H_{\a(s)\ad(s)}~,\n\\
\d\Gamma_{\b\a(s)\bd\ad(s)}&=-\frac{1}{s!}~C_{\b(\a_s}\Dd_{\bd}\D^2\bar{L}_{\a(s-1))\ad(s)}\sn\\
                            &~~-\frac{i}{(s+1)!}~\pa_{\b(\bd}\Dd_{\ad_s}L_{\a(s)\ad(s-1))}\\
                            &~~-\frac{1}{(s+1)!}~C_{\bd(\ad_s}\D_{\b}\Dd^2L_{\a(s)\ad(s-1))}\\
                            &~~-\frac{1}{(s+1)!}~C_{\bd(\ad_s}\Dd^2\D_{\b}L_{\a(s)\ad(s-1))}\\
        &+\frac{s-1}{(s+1)!}~C_{\bd(\ad_s}\Dd^{\gd}\D_{\b}\Dd_{\ad_{s-1}}L_{\a(s)|\gd|\ad(s-2))}~,
\eea
\bea{ll}
\Gamma_{\g\b\a(s)\bd\ad(s)}&=\D_{\g}\Dd_{\bd}\D_{\b}H_{\a(s)\ad(s)}~,\n\\
\d\Gamma_{\g\b\a(s)\bd\ad(s)}&=-\frac{i}{s!}~C_{\b(\a_s}\pa_{\g\bd}\D^2\bar{L}_{\a(s-1))\ad(s)}\sn\label{dG0}\\
&\hspace{-3mm}-\frac{i}{(s+1)!}~\pa_{\b(\bd}\D_{\g}\Dd_{\ad_s}L_{\a(s)\ad(s-1))}\\
 &\hspace{-3mm}+\frac{1}{(s+1)!}~C_{\bd(\ad_s}C_{\g\b}\D^2\Dd^2L_{\a(s)\ad(s-1))}\\
&\hspace{-3mm}+\frac{s-1}{(s+1)!}~C_{\bd(\ad_s}\D_{\g}\Dd^{\gd}\D_{\b}\Dd_{\ad_{s-1}}L_{\a(s)|\gd|\ad(s-2))}\\
 &\hspace{-3mm}-\frac{1}{(s+1)!}~C_{\bd(\ad_s}\D_{\g}\Dd^2\D_{\b}L_{\a(s)\ad(s-1))}
\eea
By imposing various (anti)symmetrizations of indices we can simplify the above
transformations. Also notice that the last term in \eqref{dG0} has the characteristic structure of
the transformation of a superconnection \eqref{scn}. So let us consider the following quantity:
\bea{l}
\Gamma_{\a(s+2)\ad(s-1)}=\frac{1}{(s+2)!}~\D_{(\a_{s+2}}\pa_{\a_{s+1}}{}^{\ad_{s}}H_{\a(s))\ad(s)}~,\\[2mm]
\d\Gamma_{\a(s+2)\ad(s-1)}=-\frac{i}{s(s+2)!}~\D_{(\a_{s+2}}\Dd^2\D_{\a_{s+1}}L_{\a(s))\ad(s-1)}
\label{dG1}\\
\hspace{14mm}-\frac{s-1}{(s+2)!s!}~\D_{(\a_{s+2}}\Dd_{(\ad_{s-1}}\pa_{\a_{s+1}}{}^{\gd}
L_{\a(s))|\gd|\ad(s-2))}.
\eea
Because of the second term in the above equation, $\Gamma_{\a(s+2)\ad(s-1)}$ is not quite yet a
superconnection. However, for the special case of $s=1$ (linearized supergravity:
$(2,3/2)$-supermultiplet) this term drops and
$\Gamma_{\a\b\g}=\tfrac{1}{3!}\D_{(\a}\pa_{\b}{}^{\gd}H_{\g)\gd}$ is the superconnection for linearized
supergravity. One can confirm that its $\bar{\theta}$-component is the  linearized Christoffel symbol
$\Dd_{\ad}\Gamma_{\a\b\g}\big\rvert_{W.Z.}\propto\pa_{(\a}{}^{\gd}h_{\b\g)\gd\ad}$.
As expected by \eqref{scw}, it defines a supercurvature tensor
$W_{\a\b\g}=\Dd^2\Gamma_{\a\b\g}\propto\Dd^2\D_{(\a}\pa_{\b}{}^{\gd}H_{\g)\gd}$ which is exactly
the known supergravity superfield strength~\cite{Wess:1977fn,Grimm:1977kp}, which includes the bosonic
and fermionic linearized curvature tensors.

It is now straightforward to define generalized higher spin superconnections by recursive
application of superspace derivatives:
\begin{widetext}
\bea{l}
\n
\Gamma^{(t)}_{\a(s+t+1)\ad(s-t)}=\frac{1}{(s+t+1)!}~\D_{(\a_{s+t+1}}\pa_{\a_{s+t}}{}^{\gd_1}\dots
\pa_{\a_{s+1}}{}^{\gd_t}H_{\a(s))\gd(t)\ad(s-t)}~,\sn\label{Gt}\\[3mm]
\delta\Gamma^{(t)}_{\a(s+t+1)\ad(s-t)}=-i~\frac{t}{s}\frac{1}{(s+t+1)!}~\D_{(\a_{s+t+1}}\Dd^2~\D_{\a_{s+t}}
\pa_{\a_{s+t-1}}{}^{\gd_1}\dots\pa_{\a_{s+1}}{}^{\gd_{t-1}}L_{\a(s))\gd(t-1)\ad(s-t)}\sn\label{dGt}\\[2mm]
\hspace{18ex}-\frac{s-t}{s}\frac{1}{(s+t+1)!(s-t)!}~\D_{(\a_{s+t+1}}\Dd_{(\ad_{s-t}}\pa_{\a_{s+t}}{}^{\gd_1}
\dots\pa_{\a_{s+1}}{}^{\gd_t}L_{\a(s))|\gd(t)|\ad(s-t-1))~.}
\eea
\end{widetext}
This is a hierarchy of $(s+1)$ superconnectionlike objects \`a la de Wit and Freedman~\cite{deWit:1979sib},
parametrized by the values of $t$ ($t=0,1,2,\dots,s$). Each of which is defined in terms of superspace
derivatives of the previous one, via the recursive relation
\seqn{\Gamma^{(t)}_{\a(s+t+1)\ad(s-t)}=\tfrac{1}{(s+t+1)!}~
\D_{(a_{s+t+1}}\Dd^{\gd_t}\Gamma^{(t-1)}_{\a(s+t))\gd_t\ad(s-t)}~.}
Only the top one ($t=s$) is a proper superconnection in the sense that
{\small
\bea{l}
\delta\Gamma^{(s)}_{\a(2s+1)}=
-\frac{i}{(2s+1)!}\n\\
\hspace{8mm}\times~\D_{(\a_{2s+1}}\Dd^2~\D_{\a_{2s}}
\pa_{\a_{2s-1}}{}^{\gd_1}\dots\pa_{\a_{s+1}}{}^{\gd_{s-1}}L_{\a(s))\gd(s-1)}
\eea
}
and as such it defines the higher spin supercurvature
\seqn{\label{hiW}W_{\a(2s+1)}=\Dd^2\Gamma^{(s)}_{\a(2s+1)}~.}
This is the exactly the invariant higher spin superfield strength constructed
in~\cite{Kuzenko:1993jp} and later found in~\cite{Gates:2013rka} by studying the transition
from irreducible, massive higher superspin representation to irreducible, massless massless higher
superspin representations.

Therefore, there actually exist a geometrical structure for higher spin
gauge superfields which naturally
extends the known super Yang-Mills and Supergravity cases. This hierarchy of higher spin superconnections
provides the supersymmetric extension of the bosonic and fermionic de Wit-Freedman higher spin connections.
One can check that they correspond to the $\bar{\theta}$ and $\bar{\theta}^2$ components of
$\Gamma^{(t)}_{\a(s+t+1)\ad(s-t)}$ respectively:
{\small
\bea{l}
\Dd_{\bd}\Gamma^{(t)}_{\a(s+t+1)\ad(s-t)}\big\rvert\propto\pa_{(\a_{s+t+1}}{}^{\ad_s}
\dots\pa_{\a_{s+2}}{}^{\ad_{s-t+1}}h_{\a(s+1))\b\ad(s)},\\[3mm]
\Dd^2\Gamma^{(t)}_{\a(s+t+1)\ad(s-t)}\big\rvert\propto\pa_{(\a_{s+t+1}}{}^{\ad_s}
\dots\pa_{\a_{s+2}}{}^{\ad_{s-t+1}}\psi_{\a(s+1))\ad(s)}.
\eea
}
\section{Extracting free equations of motion}\label{subsec:eom}
Ordinary free field theory requires a second (first) order equation for bosons (fermions)
which translates to four (two) spinorial superspace derivatives. However, the previous
geometrical approach to higher spin supermultiplets indicates that the only gauge invariant quantities
involve higher derivatives. Therefore, it is not clear how one can obtain reasonable free superfield
equations. The answer \footnote{Under the requirement of a local theory.} is that only if
appropriate constraints are imposed on the gauge parameter this can be achieved. This behavior
is homologous to nonsupersymmetric higher spins
where a traceless condition must be imposed and the fields are restricted to $SO(D)$ irreducible
tensors instead of $GL(D)$ tensors and thus forcing the constrained formulation.

Motivated from \eqref{hiW}, one can define its $t$-generalization
\seqn{W^{(t)}_{\a(s+t+1)\ad(s-t)}=\Dd^2\Gamma^{(t)}_{\a(s+t+1)\ad(s-t)}~.}
This is an interesting, secondary hierarchy because its top is the invariant superfield strength, but the
members of it are not superconnectionlike objects because their gauge transformation is:
{\small
\bea{l}
\d W^{(t)}_{\a(s+t+1)\ad(s-t)}=
-\frac{i(s-t)}{s(s+t+1)!(s-t)!}\n\label{dWt}\\
\hspace{5mm}\times~\Dd^2\pa_{(\a_{s+t+1}(\ad_{s-t}}\pa_{\a_{s+t}}{}^{\gd_1}\dots\pa_{\a_{s+1}}{}^{\gd_t}
L_{\a(s))|\gd(t)|\ad(s-t-1))}
\eea
}
In~\cite{Gates:2013rka} a deep relationship between the invariant superfield strength and the on shell
equations of motion was discovered. It was shown that the quantity $\D^{\a_{2s+1}}W_{\a(2s+1)}$ is
expressed as the sum of higher derivative operators acting on the free equations of motion. It is
natural to attempt this for the entire secondary hierarchy. One can show that:
\begin{widetext}
\bea{l}
\D^{\a_{s+t+1}}W^{(t)}_{\a(s+t+1)\ad(s-t)}=\frac{1}{(s+t+1)!}~
\pa_{(\a_{s+t}}{}^{\gd_t}\dots\pa_{\a_{s+1}}{}^{\gd_1}\Bigl\{\vphantom{\frac12}
    (s+t+1)~\D^{\b}\Dd^2\D_{\b}H_{\a(s))|\gd(t)|\ad(s-t)}\Bigr.\n\label{DW}\\
    \hspace{9.5cm}\Bigl. +s~\D_{\a_s}\Dd^2\D^{\b}H_{|\b|\a(s-1))|\gd(t)|\ad(s-t)}\Bigr\}\\
\hspace{4cm}-i~\frac{t}{(s+t+1)!}~\D_{(\a_{s+t}}\pa_{\a_{s+t-1}}{}^{\gd_{t-1}}\dots
\pa_{\a_{s+1}}{}^{\gd_1}\Dd^2\D^2\Dd^{\gd_t}H_{\a(s))|\gd(t)|\ad(s-t)~.}
\eea
\end{widetext}
The answer is similar, we get the sum of higher derivative operators acting on terms that have
the correct characteristics (engineering dimensions and index structures) to appear in the
equations of motions. The $t=0$ level seems appropriate in order not to have higher derivatives.
Under the gauge transformation we get:
\begin{widetext}
\bea{rl}
\d\Bigl(\D^{\a_{s+t+1}}W^{(t)}_{\a(s+t+1)\ad(s-t)}&\Bigr)=
\frac{(s-t)(s+1)}{s(s+t+1)}~\frac{1}{(s+t)!(s-t)!}~\Dd_{(\ad_{s-t}}\D^2\Dd^2\pa_{(\a_{s+t}}{}^{\gd_t}\dots
\pa_{\a_{s+1}}{}^{\gd_1}L_{\a(s))|\gd(t)|\ad(s-t-1))}\n\\[1mm]
+~&\frac{(s-t)t}{s(s+t+1)}~\frac{1}{(s+t)!(s-t)!}~\Dd^{\gd_t}\Dd^2\Dd^2\pa_{(\a_{s+t}(\ad_{s-t}}
\pa_{\a_{s+t-1}}{}^{\gd_{t-1}}\dots\pa_{\a_{s+1}}{}^{\gd_1}L_{\a(s))|\gd(t)|\ad(s-t-1))}\\[1mm]
-~&\frac{s-t}{s+t+1}~\frac{1}{(s+t)!(s-t)!}~\D_{(\a_{s+t}}\Dd_{(\ad_{s-t}}\D^{\b}\Dd^2\pa_{\a_{s+t-1}}{}^{\gd_t}\dots\pa_{\a_{s}}{}^{\gd_1}L_{|\b|\a(s-1))|\gd(t)|\ad(s-t-1))}
\eea
\end{widetext}
and for $t=0$ it simplifies
\bea{ll}
\d\Bigl(\D^{\a_{s+1}}W^{(0)}_{\a(s+1)\ad(s)}\Bigr)=&
~~\frac{1}{s!}~\Dd_{(\ad_{s}}\D^2\Dd^2 L_{\a(s)\ad(s-1))}\n\label{dDW0}\\
&\hspace{-20mm}-~\frac{s}{s+1}~\frac{1}{s!s!}~\D_{(\a_{s}}\Dd_{(\ad_{s}}\D^{\b}\Dd^2 L_{|\b|\a(s-1))\ad(s-1))}.
\eea
\subsection{Nonminimal Constraints}\label{subsec:nmc}
Based on the above transformation law, the quantity $\D^{a_{s+1}}W^{(0)}_{\a(s+1)\ad(s)}$ is
invariant if we constraint the gauge parameter $L_{\a(s)\ad(s-1)}$ as follows:
\seqn{\Dd^2 L_{\a(s)\ad(s-1)}+\D^{\a_{s+1}}\Lambda_{\a(s+1)\ad(s-1)}=0\label{nmc}}
where $\Lambda_{\a(s+1)\ad(s-1)}$ is an arbitrary superfield consistent with the
condition
$\Dd_{\bd}\D^{\a_{s+1}}\Lambda_{\a(s+1)\ad(s-1)}=0$~.
It is evident that, under this constraint, the gauge invariant
equation
\seqn{\mathcal{E}_{\a(s)\ad(s)}\propto\D^{\a_{s+1}}\Dd^2\Gamma^{(0)}_{\a(s+1)\ad(s)}=0\label{nonmineom}}
can play the role of free equation of motion for the $(s+1,s+1/2)$ supermultiplet. This equation
is geometrical in nature because it involves only superconnection $\Gamma^{(0)}$.
Using \eqref{DW}, eq. \eqref{nonmineom} can be decomposed to the following two equations:
\seqn{\label{nmeom}\D^{\b}\Dd^2\D_{\b}H_{\a(s)\ad(s)}=0~,~\Dd^2\D^{\b}H_{\b\a(s-1)\ad(s)}=0}
which are both gauge invariant due to \eqref{nmc}.
Later we will show that \eqref{nmeom} and \eqref{nmc} generate the known nonminimal description
of $(s+1,s+1/2)$ supermultiplets.
\subsection{Minimal Constraints}
Transformation \eqref{dDW0} can be written in a different way by realizing that \eqref{dH}
allows the decomposition of it to the sum of $\delta H$-dependent terms and the remainder.
The $\delta H$-terms can then be absorbed to the left-hand side of the equation. By isolating as many as
possible $\d H$-terms, the remainder will provide an alternative structure of softer constraints.
Specifically, the first term of \eqref{dDW0} can be written as:
\bea{l}
\frac{1}{s!}~\Dd_{(\ad_s}\D^2\Dd^2 L_{\a(s)\ad(s-1))}=\n\\
~-\frac{s(s+1)}{2s+1}~\frac{1}{s!}~\Dd_{(\ad_s}\D^2\Dd^{\rd}~\d H_{\a(s)|\rd|\ad(s-1))}\\
   ~+\frac{s^2}{2s+1}~\frac{1}{s!s!}~\Dd_{(\ad_s}\D_{(\a_s}\Dd^{\rd}\D^{\r}~\d
       H_{|\r|\a(s-1))|\rd|\ad(s-1))}\\
   ~-\frac{s^2}{2s+1}~\frac{1}{s!s!}~\Dd_{(\ad_s}\D_{(\a_s}\Dd^2\D^{\r}L_{|\r|\a(s-1))\ad(s-1))}\\
   ~+\frac{s(s-1)}{2s+1}~\frac{1}{s!s!}~\Dd_{(\ad_s}\D_{(\a_s}\Dd_{\ad_{s-1}}\D^{\r}\Dd^{\rd}
       L_{|\r|\a(s-1))|\rd|\ad(s-2))}~.
\eea
Similarly the second term:
\bea{l}
   \frac{1}{s!s!}~\D_{(\a_s}\Dd_{(\ad_s}\D^{\b}\Dd^2 L_{|\b|\a(s-1))\ad(s-1))}=\n\\
      ~+ \frac{s^2}{2s+1}~\frac{1}{s!}~\D_{(\a_s}\Dd^2\D^{\r}~\d H_{|\r|\a(s-1))\ad(s)}\\
   ~-\frac{s(s+1)}{2s+1}~\frac{1}{s!s!}~\D_{(\a_s}\Dd_{(\ad_s}\D^{\r}\Dd^{\rd}~\d
   H_{|\r|\a(s-1))|\rd|\ad(s-1))}\\
   ~-\frac{s(s+1)}{2s+1}~\frac{1}{s!s!}~\D_{(\a_s}\Dd_{(\ad_s}\D^2\Dd^{\rd}\bar{L}_{\a(s-1))|\rd|\ad(s-1))}\\
   ~+\frac{(s+1)(s-1)}{(2s+1)s!s!}~\D_{(\a_s}\Dd_{(\ad_s}\D_{\a_{s-1}}\Dd^{\rd}\D^{\r}
   \bar{L}_{|\r|\a(s-2))|\rd|\ad(s-1))}~.
\eea
Therefore, by considering the quantity
\bea{ll}
I_{\a(s)\ad(s)}=&\D^{\a_{s+1}}W^{(0)}_{\a(s+1)\ad(s)}\n\\
   &+\frac{s(s+1)}{2s+1}~\frac{1}{s!}~\Dd_{(\ad_s}\D^2\Dd^{\rd}H_{\a(s)|\rd|\ad(s-1))}\\
   &+\frac{s^3}{(2s+1)(s+1)}~\frac{1}{s!}~\D_{(\a_s}\Dd^2\D^{\r}H_{|\r|\a(s-1))\ad(s)}\\
   &-\frac{s^2}{2s+1}~\frac{1}{s!s!}~\Dd_{(\ad_s}\D_{(\a_s}\Dd^{\rd}\D^{\r}H_{|\r|\a(s-1))|\rd|\ad(s-1))}\\
   &-\frac{s^2}{2s+1}~\frac{1}{s!s!}~\D_{(\a_s}\Dd_{(\ad_s}\D^{\r}\Dd^{\rd}H_{|\r|\a(s-1))|\rd|\ad(s-1))}
\eea
we get the following transformation law:
\begin{widetext}
\bea{ll}
\d I_{\a(s)\ad(s)}=&
-\frac{s^2}{2s+1}~\frac{1}{s!s!}~\Dd_{(\ad_s}\D_{(\a_s}\Bigl[\Dd^2\D^{\g}L_{|\g|\a(s-1))\ad(s-1))}
-\frac{s-1}{s}~\Dd_{\ad_{s-1}}\D^{\g}\Dd^{\gd}L_{|\g|\a(s-1))|\gd|\ad(s-2))}\Bigr]\n\label{dI}\n\\
&+\frac{s^2}{2s+1}~\frac{1}{s!s!}~\D_{(\a_s}\Dd_{(\ad_s}\Bigl[\D^2\Dd^{\gd}\bar{L}_{\a(s-1)|\gd|\ad(s-1))}
-\frac{s-1}{s}~\D_{\a_{s-1}}\Dd^{\gd}\D^{\g}\bar{L}_{|\g|\a(s-2))|\gd|\ad(s-1))}\Bigr]
\eea
\end{widetext}
Gauge invariance is achieved if we constrained the gauge parameter $L_{\a(s)\ad(s-1)}$ in the
following way:
\bea{l}\n\label{mc}
s>1:\nn\\[2mm]
\D^{\g}\Dd^{\gd}L_{\g\a(s-1)\gd\ad(s-2)}+\frac{s}{s-1}~\Dd^{\gd}\D^{\g}L_{\g\a(s-1)\gd\ad(s-2)}\\
\hspace{1.5cm}+\frac{s}{(s-1)!}~\Dd_{(\ad_{s-2}}J_{\a(s-1)\ad(s-3))}=0~,\sn\\[2mm]
s=1:~\Dd^2\D^{\g}L_{\g}=0\sn
\eea
where $J_{\a(s-1)\ad(s-3)}$ is an arbitrary superfield.
These are weaker constraints that will be later shown to generate the known minimal description of the
half-integer superspin supermultiplet.
In this constraint formulation the gauge invariant equation
\seqn{\label{mineom}\mathcal{E}_{\a(s)\ad(s)}\propto I_{\a(s)\ad(s)}=0}
should be considered the free equations of motion. This equation of motion yields the following
two equations:
\bea{ll}
&\D^{\b}\Dd^2\D_{\b}H_{\a(s)\ad(s)}\n\label{meom1}\\[1mm]
&+\frac{s(s+1)}{2s+1}~\frac{1}{s!}
   \Bigl[\D_{(\a_s}\Dd^2\D^{\g}H_{|g|\a(s-1))\ad(s)}+c.c.\Bigr]\\[1mm]
&-\frac{s^2}{(2s+1)s!s!}
  \Bigl[\D_{(\a_s}\Dd_{(\ad_s}\Dd^{\gd}\D^{\g}H_{|\g|\a(s-1))|\gd|\ad(s-1))}+c.c.\Bigr]\\
&\hfill=0~,\\[2mm]
&s>1:~\Dd^{\gd}\D^{\g}\Dd^{\rd}H_{\g\a(s-1)\gd\rd\ad(s-2)}=0~~\text{or}\n\sn\label{meom2}\\[2mm]
&s=1:~\Dd^2\D^{\g}\Dd^{\gd}H_{\g\gd}=0~.\sn
\eea
An interesting observation is that this minimally constraint formulation, unlike the previous
nonminimally constraint formulation, does not have a geometrical
origin, in the sense that the equation of motion \eqref{mineom} can not be written purely in terms
of the superconnection $\Gamma^{(0)}_{\a(s+1)\ad(s)}$, but additional terms depending on the gauge
superfield had to be added.
\subsection{More Minimal Constraints}\label{subsec:mmc}
There is yet another way of constraining the gauge parameter in order to find sensible
equations. Using the approach of extracting $\d H$-terms, we can rewrite
\eqref{dDW0} alternatively as follows:
\bea{l}
\d\Bigl(\D^{\a_{s+1}}W^{(0)}_{\a(s+1)\ad(s)}\Bigr)=\n\\
~\frac{s}{s+1}~\frac{sc^*}{s(c^*+1)+1}~\frac{1}{s!}~\D_{(\a_s}\Dd^2\D^{\r}\d H_{|\r|\a(s-1))\ad(s)}\\[2mm]
-\frac{s}{s(c+1)+1}~\frac{1}{s!}~\Dd_{(\ad_s}\D^2\Dd^{\rd}\d H_{\a(s)|\rd|\ad(s-1))}\\[2mm]
+\frac{s}{s(c+1)+1}~\frac{1}{s!s!}~\Dd_{(\ad_s}\D_{(\a_s}
    \Bigl[\Dd^{\rd}\Dd^2\bar{L}_{\a(s-1))|\rd|\ad(s-1))}\Bigr.\\
    \hfill-c~\Bigl.\D^{\r}\Dd^2 L_{|\r|\a(s-1))\ad(s-1))}\Bigr]\\[2mm]
-\frac{s}{s(c^*+1)+1}~\frac{1}{s!s!}~\D_{(\a_s}\Dd_{(\ad_s}
\Bigl[\D^{\r}\Dd^2 L_{|\r|\a(s-1))\ad(s-1))}\Bigr.\\
\hfill-c^*~\Bigl.\Dd^{\rd}\D^2\bar{L}_{\a(s-1))|\rd|\ad(s-1))}\Bigr]
\eea
for arbitrary complex number $c$. However if $c$ is chosen to be a phase \footnote{After
redefinitions we consider only the cases of $c=-1,1$~.}, then we can impose the constraint
\seqn{\D^{\r}\Dd^2 L_{\r\a(s-1)\ad(s-1)}\pm~\Dd^{\rd}\D^2\bar{L}_{\a(s-1)\rd\ad(s-1)}=0\label{newmc}~.}
Under the assumption of this constraint, we conclude that the
equation
\bea{ll}
\mathcal{E}_{\a(s)\ad(s)}&=\D^{\a_{s+1}}W^{(0)}_{\a(s+1)\ad(s)}\n\\[1mm]
&-\frac{s}{s+1}~\frac{sc}{s(c+1)+1}~\frac{1}{s!}~\D_{(\a_s}\Dd^2\D^{\r} H_{|\r|\a(s-1))\ad(s)}\\[1mm]
&+\frac{s}{s(c+1)+1}~\frac{1}{s!}~\Dd_{(\ad_s}\D^2\Dd^{\rd} H_{\a(s)|\rd|\ad(s-1))}\\
&=0
\eea
with $c=-1,1$ is suitable to play the role of the equation of motion. As in the previous case, this is not a
geometrical equation because of the explicit $H$-terms. Unlike the previous cases, the above equation
yields only one gauge invariant equation of motion for $H$
\bea{ll}
0=&\D^{\b}\Dd^2\D_{\b}H_{\a(s)\ad(s)}\n\label{newmeom}\\[1mm]
&+\frac{s}{s(c+1)+1}~\frac{1}{s!}~\D_{(\a_s}\Dd^2\D^{\g}H_{|\g|\a(s-1))\ad(s)}\\[1mm]
&+\frac{s}{s(c+1)+1}~\frac{1}{s!}~\Dd_{(\ad_s}\D^2\Dd^{\gd}H_{\a(s)|\gd|\ad(s-1))}~.
\eea
The $s=1$ case is special. For $s=1$, constraint \eqref{newmc} has a solution
in superspace
\seqn{L_{\a}=
    \begin{cases}
        i\D_{\a}L+\Dd^{\ad}\Lambda_{\a\ad}~,~\text{for}~c=-1\\[2mm]
        \D_{\a}L+\D^{\ad}\Lambda_{\a\ad}~,~\text{for}~c=+1
    \end{cases}
}
where $L$ is an arbitrary real scalar ($L=\bar{L}$) and $\Lambda_{\a\ad}$ is an arbitrary vector
superfield. Therefore, the $s=1$ version of \eqref{newmeom} remains valid for the unconstrained
gauge transformation
\seqn{\d H_{\a\a}=
    \begin{cases}
\frac{1}{2}\pa_{\a\ad}L+\frac{1}{2}\left(\D^2\Lambda_{\a\ad}+\Dd^2\Lambda_{\a\ad}\right)~,~
c=-1\\[2mm]
\frac{1}{2}[\D_{\a},\Dd_{\ad}]L+\frac{1}{2}\left(\D^2\Lambda_{\a\ad}+\Dd^2\Lambda_{\a\ad}\right)~,~
c=1
    \end{cases}
}
The first one corresponds to the new-minimal~\cite{Akulov:1976ck,Sohnius:1981tp,Howe:1981et,Gates:1981tu}
description of linearized supergravity supermultiplet
and the second one to the new-new-minimal (or virial)~\cite{Buchbinder:2002gh,Gates:2003cz,Nakayama:2014kua}
description of linearized supergravity.
However, for general $s$, equation \eqref{newmeom} is only valid under the assumption of the constraint
\eqref{newmc} and corresponds to the higher spin version of the new minimal and new-new minimal
descriptions.

It is important to emphasize that all the different constrained formulations presented above based
on minimal or nonminimal constraints describe the same physical degrees of freedom, on shell. All of them
have the same invariant superfield strength $W_{\a(2s+1)}$ and their equations of motion are such
that when substituted in \eqref{DW} give the same on shell condition $\D^{\a(s+1)}W_{\a(s+1)}=0$.
\section{Nongeometrical Unconstrained Formulation: Compensators}\label{subsec:uncformulation}
It would be desirable if we could have unconstrained formulations, for the above descriptions.
An easy method of doing that is via the
introduction of so called compensator superfields.
Unfortunately, this approach is not geometrical because the
compensators are unrelated to the superconnections or the supercurvatures.
The way it works is that we lift the various constraints by introducing
appropriate compensators and assigning them transformations laws proportional to
the constraints. The compensators will modify the right-hand size of the proposed equations of motion
accordingly such that the gauge invariance of the equation is maintained under the full, unconstrained
transformation.

In particular, one can lift constraint \eqref{nmc} by introducing a fermionic compensator
$\chi_{\a(s)\ad(s-1)}$ equipped with the transformation law
\seqn{\d\chi_{\a(s)\ad(s-1)}\propto\Dd^2 L_{\a(s)\ad(s-1)}+\D^{\a_{s+1}}\Lambda_{\a(s+1)\ad(s-1)}~.}
This compensator will modify the right-hand side of equations \eqref{nmeom} according to \eqref{dDW0}
so they remain invariant for arbitrary gauge parameter $L_{\a(s)\ad(s-1)}$. This process will give
the known nonminimal description of $(s+1,s+1/2)$
supermultiplet~\cite{Kuzenko:1993jp,Gates:2013rka}. Likewise for \eqref{mc}, introduce compensators
$\chi_{\a(s-1)\ad(s-2)}$ for $s>1$ and chiral $\Phi$ for $s=1$ with transformations
\bea{ll}\n
\d\chi_{\a(s-1)\ad(s-2)}\propto &~\Dd^{\gd}\D^{\g}L_{\g\a(s-1)\gd\ad(s-2)}\sn\\[1mm]
                                &+\frac{s-1}{s}~\D^{\g}\Dd^{\gd}L_{\g\a(s-1)\gd\ad(s-2)}\\[1mm]
                                &+\frac{1}{(s-2)!}~\Dd_{(\ad_{s-2}}J_{\a(s-1)\ad(s-3))}~,\\[2mm]
    \d\Phi\propto\Dd^2\D^{\g}L_{\g}\sn
\eea
which will modify equations (\ref{meom1}),~(\ref{meom2}) according to \eqref{dI}. The result will be
identical to the minimal description of $(s+1,s+1/2)$
supermultiplet~\cite{Kuzenko:1993jp,Gates:2013rka} and for the $s=1$ case this will give the old-minimal
description of linearized supergravity. Finally, constraint \eqref{newmc} requires the introduction
of a real (imaginary) linear compensator $U_{\a(s-1)\ad(s-1)}$ with the following transformation law
\bea{ll}
\d U_{\a(s-1)\ad(s-1)}=&\D^{\r}\Dd^2L_{\r\a(s-1)\ad(s-1)}\n\label{dU}\\[1mm]
&\pm~\Dd^{\rd}\D^2\bar{L}_{\a(s-1)\rd\ad(s-1)}
\eea
however such a transformation completely eliminates the compensator \footnote{Any real
    (imaginary) linear superfield $U_{\a(s-1)\ad(s-1)}$ can be expressed in terms of an
    unconstrained superfield (prepotential) $\psi_{\a(s)\ad(s-1)}$,
    $U_{\a(s-1)\ad(s-1)}=\D^{\a_s}\Dd^2\psi_{\a(s)\ad(s-1)}\pm\Dd^{\ad_s}\D^2\bar{\psi}_{\a(s-1)\ad(s)}$.
    Based on \eqref{dU} the transformation of $\psi$ is algebraic
    $\d\psi_{\a(s)\ad(s-1)}=L_{\a(s)\ad(s-1)}+\Dd^{\b}\Xi_{\b\a(s)\ad(s-1)}$ and thus it can be set
to zero immediately.} itself, forcing back on us the constraint \eqref{newmc}. Therefore, the
nongeometrical method of compensators can not provide an unconstrained formulation for this case.
However at the constrained formulation, this is a new and consistent description of the  $(s+1,s+1/2)$
supermultiplet which generalizes the $s=1$ limit of new-minimal and virial linearized supergravity.
Of course, as we previously mentioned, for $s=1$ the constraint can be
explicitly solved in superspace introducing new unconstrained gauge parameters and thus effectively
making the formulation unconstrained without the need of introducing a compensator.
\section{Double hierarchy of de Wit-Freedman superconnections for integer superspins}\label{sec:idWFs}
Now let us consider integer superspin supermultiplets $(s+1/2,s)$ where the highest
propagating spin is a fermion. The appropriate superfield for the description of this supermultiplet
is a fermionic $(s,s-1)$-superfield tensor $\Psi_{\a(s)\ad(s-1)}$ with a transformation \footnote{For
the special case of $s=1$ the transformation takes the form $\d\Psi_{\a}=\D_{\a}K+\Dd^2\Lambda_{\a}$~.}
\begin{widetext}
\seqn{\d\Psi_{\a(s)\ad(s-1)}=\frac{1}{s!}~\D_{(\a_s}K_{\a(s-1))\ad(s-1)}
                            +\frac{1}{(s-1)!}~\Dd_{(\ad_{s-1}}\Lambda_{\a(s)\ad(s-2))}\label{dPsi}~.}
\end{widetext}
The highest rank component of this superfield is a symmetric rank $s$ spinor tensor
$\psi_{\a(s+1)\ad(s)}\propto
\tfrac{1}{(s+1)!s!}~{\small[\D_{(\a_{s+1}},\Dd_{(\ad_s}]\Psi_{\a(s))\ad(s-1))}\big\rvert}$
and the transformation law \eqref{dPsi} is the most general which will give the proper gauge
transformation to the above gauge field ($\delta\psi_{\a(s+1)\ad(s)}\propto
\pa_{(\a_{s+1}(\ad_s}\zeta_{\a(s))\ad(s-1))}$).

Immediately, one can observe a very important qualitative difference with the half-integer
case. There are two independent symmetries. One is parametrized by gauge parameter
$K_{\a(s-1)\ad(s-1)}$ and the other by gauge parameter $\Lambda_{\a(s)\ad(s-2)}$. For the half
integer superspin case, the reality condition of superfield $H_{\a(s)\ad(s)}$ forced a relation
between the two parameters by complex conjugation and thus collapsed the two symmetries into one.
The implications of this are that we can construct two types of superconnections:
$K$-superconnections and $\Lambda$-superconnections. Each type will have it's own hierarchy,
and supercurvatures.
\subsection{$\L$-superconnections}\label{Lsc}
Gauge parameter $\L_{\a(s)\ad(s-2)}$ appears in \eqref{dPsi} exactly the same way as
$L_{\a(s)\ad(s-1)}$ appears in \eqref{dH}.
Therefore, we can immediately inherit the results of
section \ref{sec:hidWFs}.
There is a hierarchy of $s$ $\L$-superconnectionlike objects parametrized by $t$
($t=0,1,2,\dots,s-1$)
\bea{ll}
\Gamma^{(t)}_{\a(s+t+1)\ad(s-t-1)}&=\frac{1}{(s+t+1)!}\n\label{pGt}\\
&\hspace{-1.5cm}\times~\D_{(\a_{s+t+1}}\pa_{\a_{s+t}}{}^{\gd_1}\dots
\pa_{\a_{s+1}}{}^{\gd_t}\Psi_{\a(s))\gd(t)\ad(s-t-1)}
\eea
which satisfy the recursive relation
\bea{l}
\Gamma^{(t)}_{\a(s+t+1)\ad(s-t-1)}\propto
\D_{(a_{s+t+1}}\Dd^{\gd_t}\Gamma^{(t-1)}_{\a(s+t))\gd_t\ad(s-t-1)}~~~\n
\eea
and have the following gauge transformation law
\begin{widetext}
\bea{l}
\delta\Gamma^{(t)}_{\a(s+t+1)\ad(s-t-1)}=i~\frac{t}{s-1}\frac{1}{(s+t+1)!}~\D_{(\a_{s+t+1}}\Dd^2~\D_{\a_{s+t}}
\pa_{\a_{s+t-1}}{}^{\gd_1}\dots\pa_{\a_{s+1}}{}^{\gd_{t-1}}\L_{\a(s))\gd(t-1)\ad(s-t-1)}\n\label{pdGt}\\[2mm]
\hspace{14ex}+\frac{s-t-1}{s}\frac{1}{(s+t+1)!(s-t-1)!}~\D_{(\a_{s+t+1}}\Dd_{(\ad_{s-t-1}}
\pa_{\a_{s+t}}{}^{\gd_1}\dots\pa_{\a_{s+1}}{}^{\gd_t}\L_{\a(s))|\gd(t)|\ad(s-t-2))~.}
\eea
\end{widetext}
The top member $\Gamma^{(s-1)}_{\a(2s)}$ is a proper superconnection and defines
an invariant supercurvature.
\seqn{W_{\a(2s)}=\Dd^2\Gamma^{(s-1)}_{\a(2s)}~.}
This is identical to the invariant superfield strength constructed
in~\cite{Kuzenko:1993jq} and later in~\cite{Gates:2013rka}.
\subsection{$K$-superconnections}\label{subsec:Ksc}
Gauge parameter $K_{\a(s-1)\ad(s-1)}$ appears in \eqref{dPsi} exactly the same way as
$\bar{L}_{\a(s-1)\ad(s)}$ in \eqref{dH}. Hence, if we use $\bar{\Psi}_{\a(s-1)\ad(s)}$ instead of
$\Psi_{\a(s)\ad(s-1)}$ in the construction of superconnections we can also use the results of
section \ref{sec:hidWFs}.
There is a hierarchy of ($s+1$) $K$-superconnection like objects
\bea{ll}
\Delta^{(t)}_{\a(s+t)\ad(s-t)}=&\frac{1}{(s+t)!}\n\label{Dt}\\
                               &\hspace{-1.3cm}\times~\D_{(\a_{s+t}}\pa_{\a_{s+t-1}}{}^{\gd_1}\dots
\pa_{\a_{s}}{}^{\gd_t}\bar{\Psi}_{\a(s-1)))\gd(t)\ad(s-t)}
\eea
which is not related to the hierarchy in \eqref{pGt} by complex conjugation and must be studied
independently. Their gauge transformation is
\begin{widetext}
\bea{ll}
\delta\Delta^{(t)}_{\a(s+t)\ad(s-t)}=&~\frac{t}{s}\frac{i}{(s+t)!}~\D_{(\a_{s+t}}\Dd^2~\D_{\a_{s+t-1}}
\pa_{\a_{s+t-2}}{}^{\gd_1}\dots\pa_{\a_{s}}{}^{\gd_{t-1}}\bar{K}_{\a(s-1))\gd(t-1)\ad(s-t)}\n\label{dDt}\\[2mm]
&+\frac{s-t}{s}\frac{1}{(s+t)!(s-t)!}~\D_{(\a_{s+t}}\Dd_{(\ad_{s-t}}
\pa_{\a_{s+t-1}}{}^{\gd_1}\dots\pa_{\a_{s}}{}^{\gd_t}\bar{K}_{\a(s-1))|\gd(t)|\ad(s-t-1))}~.
\eea
\end{widetext}
The top member of this hierarchy, $\Delta^{(s)}_{\a(2s)}$ is a superconnection and it defines the
following supercurvature
\seqn{Z_{\a(2s)}=\Dd^2\Delta^{(s)}_{\a(2s)}~.}
\subsection{Free equations of motion}
Again, the invariant tensors involve higher derivatives. Hence, the extraction of proper
equations of motion for the $(s+1/2,s)$ supermultiplet must relay on constraining the gauge
parameters, to decouple the higher order members in at least one of the two hierarchies.
The relevant quantities to investigate are:
$\Dd^{\ad_s}\Delta^{(0)}_{\a(s)\ad(s)}$,~$\Dd^{\ad_s}\bar{\Delta}^{(0)}_{\a(s)\ad(s)}$ and
$\D^{\a_s}\Gamma^{(0)}_{\a(s+1)\ad(s-1)}$~.
For, $s>1$, we find:
\bea{ll}
&\d\Bigl(\Dd^{\ad_s}\Delta^{(0)}_{\a(s)\ad(s)}\Bigr)=-\frac{1}{s!}\D_{(\a_s}\Dd^2\bar{K}_{\a(s-1))\ad(s-1)}
\\[1mm]
&\hspace{2.5cm}-\frac{1}{s!}\Dd^2\D_{(\a_s}\bar{K}_{\a(s-1))\ad(s-1)}\\[1mm]
&\hspace{2.5cm}+\frac{s-1}{s!s!}\Dd_{(\ad_{s-1}}\D_{(\a_s}\Dd^{\rd}\bar{K}_{\a(s-1))|\rd|\ad(s-2))}~,\\[3mm]
&\d\Bigl(\Dd^{\ad_s}\bar{\Delta}^{(0)}_{\a(s)\ad(s)}\Bigr)=
    -\frac{s+1}{s}~\frac{1}{s!}~\Dd^2\D_{(\a_s}K_{\a(s-1))\ad(s-1)~,}\\[3mm]
&\d\Bigl(\D^{\a_{s+1}}\Gamma^{(0)}_{\a(s+1)\ad(s-1)}\Bigr)=\frac{s(s+2)}{(s+1)!}
    \D^2\Dd_{(\ad_{s-1}}\L_{\a(s)\ad(s-2))}~.
\eea
By constraining gauge parameter $K_{\a(s-1)\ad(s-1)}$ in the following way
\bea{l}
\bar{K}_{\a(s-1)\ad(s-1)}\pm K_{\a(s-1)\ad(s-1)}=0~,\n\label{pLc}\\ 
    \D^{\b}K_{\b\a(s-2)\ad(s-1)}=0\Rightarrow K_{\a(s-1)\ad(s-1)}=\D^{\a_s}L_{\a(s)\ad(s-1)}
\eea
we get the following gauge invariant equation of motion
\seqn{\mathcal{E}_{\a(s)\ad(s-1)}\propto
\Dd^{\ad_{s+1}}\Bigl(\frac{s+1}{s}\Delta^{(0)}_{\a(s)\ad(s)}\pm\bar{\Delta}^{(0)}_{\a(s)\ad(s)}\Bigr)~.}
The equation of motion is expressed purely in terms of superconnection $\Delta^{(0)}$ and in the
constrained formulation it
produces the following two gauge invariant equations for superfield $\Psi_{\a(s)\ad(s-1)}$
\bea{l}\n\label{peom}
\frac{1}{s!}~\Dd^{\ad_s}\D_{(\a_s}\bar{\Psi}_{\a(s-1)\ad(s)}\mp\Dd^2\Psi_{\a(s)\ad(s-1)}=0~,\sn\\[2mm]
\D^{\a_s}\Dd^2\Psi_{\a(s)\ad(s-1)}\pm\Dd^{\ad_s}\D^2\bar{\Psi}_{\a(s-1)\ad(s)}=0~.\sn
\eea
Constraint \eqref{pLc} is lifted via a
real (imaginary) bosonic compensator $V_{\a(s-1)\ad(s-1)}$, with transformation
\bea{l}
\d V_{\a(s-1)\ad(s-1)}\propto\D^{\a_s}L_{\a(s)\ad(s-1)}\pm\Dd^{\ad_s}\bar{L}_{\a(s-1)\ad(s)}~.~~~~\n
\eea
The compensator will modify the right-hand side of \eqref{peom} accordingly so they
remain invariant under the full symmetry without constraint \eqref{pLc}. This
unconstrained formulation gives precisely the integer superspin description
of~\cite{Kuzenko:1993jq,Gates:2013rka}.

However, for the special case of $s=1$ there is an alternative constraint that one can impose.
In that case, we find:
\bea{rl}
\d\Bigl(\Dd^{\ad}\Delta^{(0)}_{\a\ad}\Bigr)=
&-\frac12~\D_{\a}\Dd^{\ad}\d\bar{\Psi}_{\ad}+\frac12~\D^2\Dd^{\ad}\D_{\a}\bar{\L}_{\ad}-\Dd^2\D_{\a}
\bar{K}~,\\[3mm]
\d\Bigl(\Dd^{\ad}\bar{\Delta}^{(0)}_{\a\ad}\Bigr)=&-2~\Dd^2\D_{\a}K~,~~
\d\Bigl(\D^{\b}\Gamma^{(0)}_{\b\a)}\Bigr)=~\frac{3}{2}~\D^2\Dd^2\L_{\a}~.
\eea
Therefore, we get
\begin{widetext}
\bea{ll}
\d\Bigl(\Dd^{\ad}\Delta^{(0)}_{\a\ad}\pm\frac12\Dd^{\ad}\bar{\Delta}^{(0)}_{\a\ad}
\pm\frac13\D^{\b}\Gamma^{(0)}_{\b\a}\Bigr)=
-\frac12\D_{\a}\Dd^{\ad}\d\bar{\Psi}_{\ad}&-\Dd^2\D_{\a}\Bigl\{\bar{K}\pm K\Bigr\}
+\frac12\Bigl\{\D^2\Dd^{\ad}\D_{\a}\bar{\L}_{\ad}\pm\D^2\Dd^2\L_{\a}\Bigr\}
\eea
\end{widetext}
Under the constraints
\bea{l}\n\label{ps1c}
\bar{K}\pm~K=0~\Rightarrow~K=
\begin{cases}
    i~L~,~\text{for}~+\\
    L~,~\text{for}~-
\end{cases}\sn
\eea
and
\bea{l}
\D^2\Dd^{\ad}\D_{\a}\bar{\L}_{\ad}\pm\D^2\Dd^2\L_{\a}=0\Rightarrow\L_{\a}=
\begin{cases}
    \D_{\a}\L~,~\text{for}~+\\
    i\D_{\a}\L~,~\text{for}~-\\
\end{cases}\sn
\eea
where $L,\L$ are arbitrary real scalar superfields,
we get the following equation of motion:
\seqn{\mathcal{E}_{\a(s)\ad(s-1)}\propto\Dd^{\ad}\Delta^{(0)}_{\a\ad}
                    \pm\frac12~\Dd^{\ad}\bar{\Delta}^{(0)}_{\a\ad}
                    \pm\frac13~\D^{\b}\Gamma^{(0)}_{\b\a}
                    +\frac12\D_{\a}\Dd^{\ad}\bar{\Psi}_{\ad}~.}
This gives an alternative to (\ref{peom}) equation for $\Psi_{\a}$
\seqn{\Dd^{\ad}\D_{\a}\bar{\Psi}_{\ad}\mp\Dd^2\Psi_{\a}
+\frac12\D_{\a}\Dd^{\ad}\bar{\Psi}_{\ad}\pm\frac12\D^2\Psi_{\a}=0~.}
Constraints \eqref{ps1c}, as indicated, are solved explicitly in superspace in terms of new
unconstrained parameters $L,\L$ thus making this description automatically unconstrained without
requiring a compensator. This equation of motion for superfield $\Psi_{\a}$ corresponds
to the description in~\cite{Ogievetsky:1975vk}.
\section{Conclusions and Discussion}\label{sec:CAD}
The description of free, manifestly supersymmetric, higher spin supermultiplets and some of their
interactions found in literature have no geometrical interpretation. The corresponding superspace actions or
equations of motion have been determined on the base of some ansatz
and gauge invariance had to be checked. Rewriting them in terms of
geometrical objects like connections and curvatures is not obvious and far from trivial if it can be done.

In this work we focus towards a more geometrical formulation of free higher superspins as described in
~\cite{Kuzenko:1993jp,Kuzenko:1993jq} and later in~\cite{Gates:2013rka}.
Starting from generic transformations of bosonic and fermionic higher spin gauge superfields, we find an
underlying geometrical structure based on the notion of higher spin superconnections. These superconnections
are seated on top of a hierarchy of superconnectionlike objects which are recursively defined by the action
of supersymmetric covariant derivatives. At the component level they  include
the de Wit-Freedman connections. Specifically, for half-integer superspin supermultiplets, we find an
$(s+1)$-hierarchy and for integer superspin supermultiplets we find two independent hierarchies with
$s$ and $s+1$ members respectively.

The top superconnections define corresponding higher spin supercurvatures which
involve higher derivatives and match the known higher spin superfield strengths. These are the only
gauge invariant objects, which makes the identification of proper superspace equations of motion unclear.
An answer is reducing the symmetry group by imposing constraints on the gauge
parameters such that the higher derivative members of the hierarchy decouple.
In contrast to nonsupersymmetric higher spins, we find several different ways of decoupling
which lead to different classes of constraints. All of these constraints, with one exception, generate
all known descriptions of higher and lower gauge supermultiplets. This was demonstrated
nongeometrically by introducing compensators and compering with known theories. For the exception,
the constrained formulation provides a new description of the  half-integer supermultiplet which mimics
the new-minimal~\cite{Akulov:1976ck,Sohnius:1981tp,Howe:1981et,Gates:1981tu} and
virial~\cite{Buchbinder:2002gh,Gates:2003cz,Nakayama:2014kua} desciption of linearized supergravities.

Furthermore, in the constrained formulation we find that for a few cases the equations of motion
are expressed purely in terms of the superconnections and thus having directly a gerometrical interpretation.
Whereas for the rest this is not possible because terms that depend on the gauge superfield had to be added.
In this sense we label theories as geometrical and nongeometrical. Out of all possible
descriptions of $(s+1,s+1/2)$ supermultiplet only one of them is geometrical and
the same holds true for $(s+1,s+1/2)$ supermultiplet.

We hope that this geometrical structure suggested by free higher spin supermultiplets can play a role in
describing consistent and nontrivial interactions in superspace. For nonsupersymmetric
theories this has been demonstrated in the framelike formulation, where the de Wit-Freedman connections are
related to extra, auxiliary higher spin connections appearing in the theory. At the free field level these
extra auxiliary fields can be decoupled~\cite{Vasiliev:1986td} precisely offering second order field equations,
however they are required for the construction of several types of nontrivial
interactions~\cite{Fradkin:1986qy,Vasiliev:2001wa,Smirnov:2013kba,Khabarov:2020bgr}.
Furthermore in~\cite{Engquist:2007yk} it was demonstrated that the construction of a covariant theory for the
higher spin algebra requires the presence of additional higher spin connections which can be expressed in
terms of the de Wit-Freedman connections.

Additionally we would like to investigate whether alternative methods of lifting the various constraints
can exist in superspace that preserve the geometrical character of the approach presented here.
For nonsupersymmetric theories such unconstrained
formulations have been found by relaxing locality~\cite{Francia:2002aa,Francia:2002pt}
or exploiting generalized versions of the Poincar\'e lemma~\cite{Damour:1987vm,Bekaert:2003az,Bekaert:2006ix}
It would also be interesting to investigate whether this
geometrical structure holds in AdS superspace. In standard field theory this has been
demonstrated in~\cite{Manvelyan:2007hv}.
\section*{Acknowledgements}
The authors thank S.~M.~Kuzenko for correspondence.
The work of I.~L.~B. was partially supported by Ministry of Science and
High Education of Russian Federation, project No FEWF-2020-003.
Also, he is grateful to RFBR grant, project No 18-02-00153 for
partial support. The research of S.~J.~G. and K.~K. is supported
in part by the endowment of the Ford Foundation Professorship of
Physics at Brown University and they gratefully acknowledge the
support of the Brown Theoretical Physics Center.

\bibliographystyle{apsrev4-2}
\bibliography{references}

\end{document}